\documentclass[aps,prl,twocolumn,nopacs,superscriptaddress]{revtex4}
\usepackage{amssymb}
\usepackage{color}
\usepackage{graphicx}
\usepackage{subfigure}
\usepackage{epsfig}
\usepackage{amsmath}
\usepackage{subfigure}
\usepackage{hyperref}

\newenvironment{proof}[1][Proof]{\noindent\textit{#1:} }{}

\begin{document}

\title{ Quantum state majorization at the output of bosonic Gaussian
channels }
\author{A. Mari}
\email{andrea.mari@sns.it}
\author{V. Giovannetti}
\affiliation{NEST, Scuola Normale Superiore and Istituto Nanoscienze-CNR, I-56127 Pisa,
Italy.}
\author{A. S. Holevo}
\affiliation{Steklov Mathematical Institute, 119991 Moscow, Russia}
\affiliation{National Research University Higher School of Economics (HSE)}

\begin{abstract}
Quantum communication theory explores the
implications of quantum mechanics to the tasks of information transmission.
Many physical channels can be formally described as quantum
Gaussian operations acting on bosonic quantum states.
Depending on the input state and on the quality of the channel, the output suffers certain amount of noise.
For a long time it has been conjectured, but never proved, that output states of Gaussian channels corresponding to coherent input signals
are the less noisy ones (in the sense of a majorization criterion).
In this work we prove this conjecture.
Specifically we show that every output state of a phase insensitive Gaussian
channel is majorized by the output  state corresponding to a coherent input.
The proof is based on the optimality of coherent states for
the minimization of strictly concave output functionals. Moreover we show
that coherent states are the unique optimizers.
\end{abstract}

\maketitle

\noindent{\bf \large Introduction}\\
Design and analysis of the optimal protocols for
processing, storing and transmitting information is the subject of the fundamental research field of information theory pioneered in the last century by  C. E. Shannon \cite{Shannon}.
In reality, information needs necessarily to be recorded onto a physical medium and
transmitted via a physical channel. Therefore, in addition to information theory, communication protocols should obey
the laws of physics. The progress in microminiaturization of data-processing systems leads to use of information carriers that cannot be described by
classical theory and behave according to quantum mechanics ({\it e.g.} photons, electrons, atoms, {\it etc.}). The task
of quantum information and communication theory is to study the laws of information transmission and processing in the quantum mechanical systems \cite{qinfo, qcomm,channels,hw}.

\begin{figure}[t]
\includegraphics[width=0.8 \columnwidth] {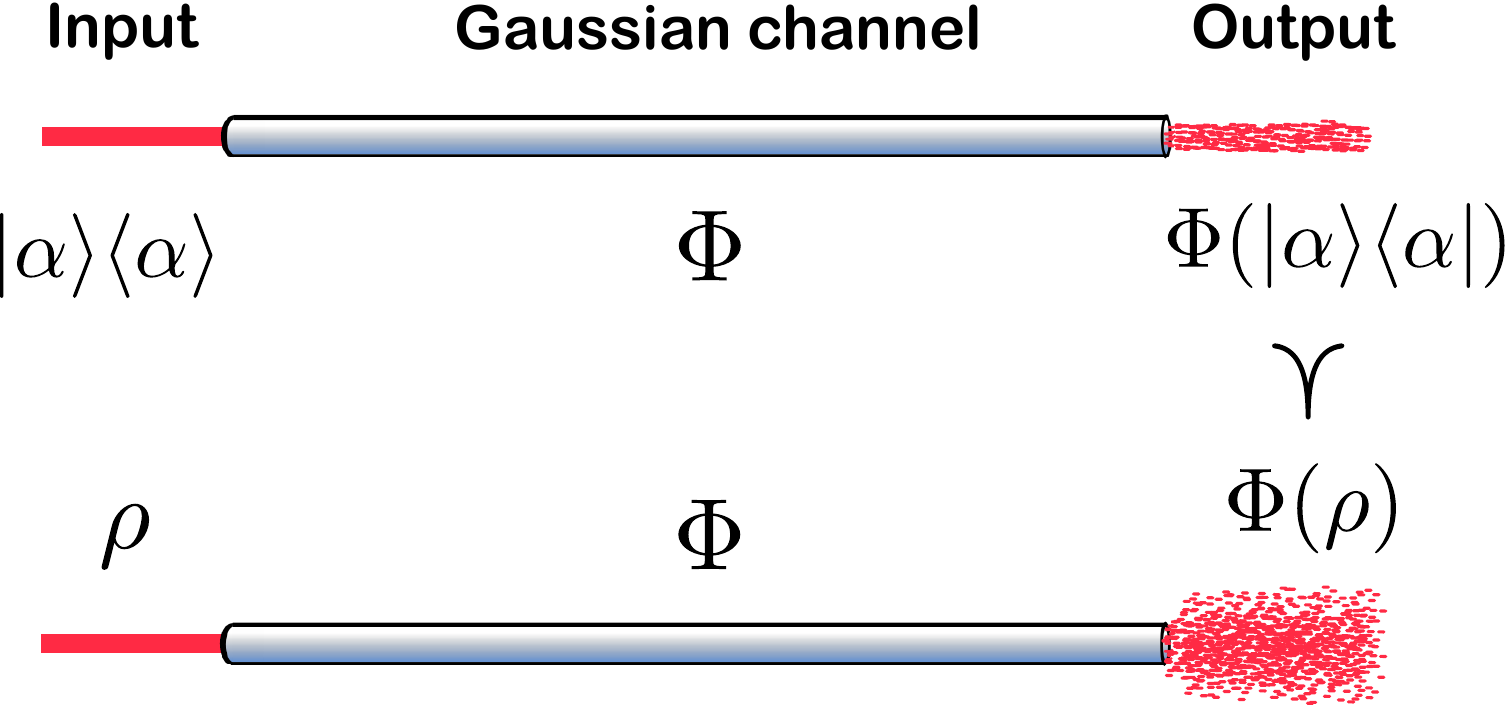}
\caption{Graphical representation of the majorization conjecture (B). A coherent state $|\alpha \rangle \langle \alpha |$ and an arbitrary state $\rho$ are both transmitted through the same phase-insensitive Gaussian channel $\Phi$. The respective output states always satisfy the majorization relation $\Phi(|\alpha \rangle \langle \alpha |) \succ \Phi( \rho)$. This means that coherent input states produce less ``noise'' at the output of the communication channel.} \label{noise}
\end{figure}

 A large part of quantum communication theory is devoted to the transmission of electromagnetic radiation via bosonic Gaussian channels \cite{channels,capacity,gauss2,gauss3}. The latter are
formally defined as completely positive and trace preserving operations mapping
Gaussian input states into Gaussian output states. The most relevant channels are also invariant under
phase space rotations and are called {phase-insensitive}. For example,
the transmission of optical quantum states through realistic physical
devices \cite{qcomm} (like \textit{e.g.} optical fibers, free space
communication lines, dielectric media, \textit{etc.}) can be described by
phase-insensitive Gaussian channels.

In the spirit of classical communication theory \cite{Shannon}, one may ask
what is the minimum amount of ``disorder''
achievable at the output of a Gaussian channel. For quantum systems there are two main
figures of merit which can be used to quantify the idea of disorder \cite{petz, nielsen, concavity,strict}: the {von Neumann
entropy} and the concept of majorization. The entropy of a state $
\rho $ is defined as $S(\rho )=-\mathrm{Tr}[\rho \log (\rho )]$ and one can
say that a state $\rho _{1}$ is more disordered than $\rho _{2}$ if $S(\rho
_{1})>S(\rho _{2})$. A different (and stronger) way of saying that $\rho _{1}
$ is more disordered than $\rho _{2}$ is the following:
\begin{equation}
\sum_{j=1}^{k}\lambda _{j}^{\rho _{1}}\leq \sum_{j=1}^{k}\lambda _{j}^{\rho
_{2}},\quad \forall k\geq 1,  \label{majorization}
\end{equation}%
where the vectors $\boldsymbol{\lambda}^{\rho_1}$ and $\boldsymbol{\lambda}^{\rho_2}$ consist of the eigenvalues of the
respective states arranged in decreasing order. If the condition (\ref%
{majorization}) is satisfied then one says that
$\rho _{2}$ majorizes
 $\rho _{1}$ and this is usually indicated by the expression $\rho
_{2}\succ \rho _{1}$. The previous definition has a very intuitive
operational interpretation since it can be shown that $\rho _{2}\succ \rho
_{1}$ if and only if $\rho _{1}$ can be obtained from $\rho _{2}$ by a
proper convex combination of unitary operations \cite{nielsen,petz,concavity,strict}. These considerations extend also to the infinite dimensional case \cite{kaftal} relevant for the quantum description of electromagnetic modes.

According to the previous ideas of disorder it was conjectured \cite{conj1} that for a phase-insensitive bosonic Gaussian channel:

\begin{quote}
(A) the minimum output entropy is achieved by coherent input states,
\end{quote}
and

\begin{quote}
(B)  the output states resulting from coherent input states majorize all
other output states.
\end{quote}
A graphical representation of the last property is given in Fig.  \ref{noise}.
Both conjectures have
broad implications in many research areas like classical
and quantum optics, telecommunication engineering, mathematical and
statistical physics and for this reason they attracted the attention of many scientists.
In particular, the validity of (A) and (B) has a number of important corollaries and relations ranging
from entanglement theory \cite{eof1,eof2,eof3,vittorio2}, channel capacities \cite%
{capacity,conj1,gp,konig1,konig2,natphot,vittorio2}, entropic inequalities \cite%
{evid1,evid3,konig1,konig2} to quantum discord \cite{discord1,discord2}.

In the last decade, many analytical and numerical evidences supporting both
conjectures were presented \cite%
{conj1,conj2,natphot,gp,evid1,evid2,evid3,evid4,evid5,evid7,konig1,konig2}
but a general proof was missing. Only very recently the first one (A) was
finally proved \cite{vittorio2,vittorio} under the assumption of a finite mean energy.
In this work we prove the second conjecture (B) and highlight some of its implications.
Moreover it is easy to show that $\rho _{2}\succ \rho _{1}$ implies $S(\rho
_{1})\geq S(\rho _{2})$, therefore the statement (B) is stronger than (A)
 and the result presented in this work can also be seen as a
proof of the minimum output entropy conjecture, without any energy
constraint. Thus both gaps in the theory are now definitely closed.
\\

\noindent{\bf \large Results}\\
\noindent{\bf Gaussian channels.}
Every quantum channel \cite{channels,NandC} can be described as a global unitary operation applied to the tensor product of the state of the system $\rho_S$ and the state of an appropriate environment  $\rho_E$:
\begin{equation}\label{channel}
\Phi (\rho_S)={\rm Tr}_E  U (\rho_{S} \otimes \rho_E) U^\dag .
\end{equation}

Single-mode phase insensitive channels \cite{natphot} can be classified in three main classes $\mathcal E_\eta^N$,  $\mathcal N_n$ and $\mathcal A_\kappa^N$.
Physically, $\mathcal E_\eta^N$ represents a thermal channel which can be realized by a beamsplitter of
transmissivity $\eta$ mixing the input signal with a thermal state $\rho_E$ with mean photon number $N$:
\begin{equation}
 U^\dag a U=  \sqrt{\eta} a  + \sqrt{1-\eta} a_E ,  \label{beamsplitter}
\end{equation}
where $a$ and $a_E$ are the annihilation operators of the system and of the environment respectively.
Then, $\mathcal N_n$ is the classical additive noise channel where the input state is displaced according to a random Gaussian
distribution of variance $n$ and, finally, $\mathcal A_\kappa^N$ is the quantum amplifier
where the state of the environment is in a thermal state $\rho_E$:
\begin{equation}
 U^\dag a U=  \sqrt{\kappa} a  + \sqrt{\kappa-1} a_E^\dag .  \label{first}
\end{equation}
More precisely, these channels can be defined according to their action on the quantum characteristic function $\chi(\mu):=\mathrm{Tr}[\rho e^{\mu a^\dag-\bar{\mu} a}]$ in the following way \cite{gauss3}:
\begin{eqnarray}
\chi(\mu)&\xrightarrow{\mathcal E_\eta^N}& \chi(\sqrt{\eta}\mu)e^{-(1-\eta)(N+1/2)|\mu|^2},\quad \eta \in (0,1),\\
\chi(\mu)&\xrightarrow{\mathcal N_n}& \chi(\mu)e^{-n|\mu|^2}, \quad n>0 \\
\chi(\mu)&\xrightarrow{\mathcal A_\kappa^N}& \chi(\sqrt{\kappa}\mu)e^{-(\kappa-1)(N+1/2)|\mu|^2}, \quad \kappa > 1. \label{amp}
\end{eqnarray}

Any of the previous phase insensitive channels, which we denote by the symbol $\Phi$, can always be decomposed  \cite{gp,vittorio} into a pure-loss
channel followed by a quantum-limited amplifier:
\begin{equation}\label{decomp}
 \Phi = \mathcal A_\kappa^0 \circ \mathcal E_\eta^0,
\end{equation}
for appropriate values of $\kappa$ and $\eta$.
In the following we will make use of this decomposition and for simplicity we will use the symbols $\mathcal A_\kappa$ and $\mathcal E_\eta$ for indicating the respective quantum-limited channels with $N=0$.\\


\noindent {\bf Complementary channels.}
 One can associate to every channel (\ref{channel}) the respective complementary channel $\tilde \Phi$ defined as
 \begin{equation}
\tilde \Phi (\rho_{S})={\rm Tr}_{S}  U (\rho_{S} \otimes \rho_E) U^\dag ,
\end{equation}
and physically representing the flow of information from the input state to the environment \cite{channels, gauss3}.
An important property of complementary channels is that, whenever the system and the environment are in a pure state, the nonzero spectra of the output  states
$\Phi (|\psi_S \rangle \langle \psi_S |)$ and $\tilde \Phi (|\psi_S \rangle \langle \psi_S |)$ are equal. This is a simple consequence of the
Schmidt decomposition of the global pure state (for an explicit proof see \cite{NandC}).

In the following the complementary channel of the quantum limited amplifier  will play an important role.
In this case $\rho_E$ is the vacuum and $U$ is the two-mode squeezing operation \cite{gauss3} acting in the Heisenberg picture as  (\ref{first}) and
\begin{equation}
 U^\dag a_E U=   \sqrt{\kappa -1} a^\dag  + \sqrt{\kappa} a_E  .  \label{second}
\end{equation}
From Eq.\ (\ref{first}), tracing out the environment, one obtains $\mathcal A_k$ defined in (\ref{amp}) with $N=0$.
From Eq.\ (\ref{second}) instead, tracing out the system, we get the complementary channel $\tilde {\mathcal A}_k$ acting on the characteristic
function as
\begin{equation}
\chi(\mu) \xrightarrow{\tilde{\mathcal A}_\kappa} \chi( -\sqrt{\kappa-1}\bar \mu)e^{-\frac{\kappa}{2} |\mu|^2} . \label{ampc}
\end{equation}
Importantly, the complementary channel does not have the same structure as the amplifier given in Eq.\ (\ref{amp}), since the complex variable $\mu$ appears conjugated in the RHS of (\ref{ampc}).
In quantum optics this effect is known as  phase conjugation or  time reversal and corresponds to the positive (but not completely positive) map
\begin{equation}
\chi(\mu) \xrightarrow{T} \chi( -\bar \mu) , \label{T}
\end{equation}
which at the level of density operators behaves as transposition $T(\rho)=\rho^{\top}$ in the Fock basis and therefore it preserves the eigenvalues.
This means that each time we are interested in spectral properties of the output state (as in the proof of Lemma 1),  we can neglect the effect of the phase conjugation operator $T$. \\


\noindent{\bf Minimization of strictly concave  functionals. }
Before giving the proof of the majorization conjecture we consider an important minimization problem.

Let $F:\mathcal  H \rightarrow \mathbb R$ be a unitary invariant and strictly concave
functional acting on the infinite-dimensional Hilbert space $\mathcal H$ of density matrices of a single
bosonic mode. We assume that $F$ can take values in $[0,+\infty]$, having in mind applications to the von Neumann entropy.
 Unitary invariance means that $F(U \rho U^\dag)=F( \rho )$ for every unitary matrix  $U$,
while  strict concavity means that
\begin{equation}
F(p \rho_1 +(1-p) \rho_2) \ge p F(\rho_1) +(1-p) F(\rho_2), \; \;p \in (0,1),
\end{equation}
and the equality is obtained only for $\rho_1=\rho_2$.
The problem that we want to address is the minimization of such functionals at the output of a phase-insensitive channel, where the optimization is performed over all possible input states:
\begin{equation}
\min_{\rho} F(\Phi(\rho)).
\end{equation}
An important case is when the functional is replaced by the von Neumann entropy  $F(\rho)=S(\rho)= -\mathrm{Tr}[\rho\log(\rho)]$, and the minimization problem reduces to the minimum output entropy conjecture (A) \cite{conj1,conj2}.
We recall that this conjecture  claims that the minimum is achieved by input coherent states of the form
\begin{equation}
|\alpha\rangle =e^{\alpha a^\dag-\bar\alpha a} |0\rangle, \quad \alpha \in \mathbb C,
\end{equation}
 and was recently proved \cite{vittorio}. With the next lemma, we are going to show that this extremal property of coherent states is more general and can be applied to every functional of the kind that we have previously introduced. \\

\noindent {\bf Lemma 1 }
{ Let $\Phi$ be a phase-insensitive bosonic channel. Then, for every nonnegative unitary invariant and strictly concave functional $F$ and for every quantum state $\rho$, we have
\begin{equation}\label{eqlemma}
F(\Phi(\rho))\ge F(\Phi(| \alpha \rangle \langle \alpha|)),  \qquad  \forall \alpha \in \mathbb C,
\end{equation}
where $| \alpha \rangle$ is any coherent state. Moreover the equality is achieved only if $\rho$ is a coherent state.} \\

\begin{proof}
For a pure loss channel $\mathcal E_\eta$, the proof is simple. Indeed coherent states are mapped to pure coherent states under the action of $\mathcal E_\eta$. Since $F$ is concave and unitary invariant, when applied to pure states it necessarily achieves its minimum.  So, in this case, Eq. (\ref{eqlemma}) is satisfied (the uniqueness property of coherent states is considered in the last part of this proof and is a consequence of Lemma 2 of section Methods).

For a general phase-insensitive channel $\Phi$ we can use the decomposition of Eq.\ (\ref{decomp}). A direct consequence of this decomposition is that we just need to prove the lemma for the minimal noise amplification channel $\Phi=\mathcal A_\kappa$ since coherent states remain coherent after a beam splitter.
Let $\tilde{\mathcal A_\kappa}$ be the  conjugate channel of $\mathcal A_\kappa$. Again, $\tilde{\mathcal A_\kappa}$  can be itself decomposed according to the structure of Eq.\ (\ref{decomp}). Indeed, from a direct application of Eq.s (\ref{amp}, \ref{ampc} ,\ref{T})  one can verify that
$$\tilde{\mathcal A_\kappa}= T  \circ \mathcal A_\kappa \circ   \mathcal E_{\eta},$$
where $T$ is the phase conjugation operator and  $\eta=1-1/\kappa$.

Let $\mathcal K$ be the set of all pure input states minimizing the functional $F$ at the output of the channel $\mathcal A_\kappa$.
We need to show that
$\mathcal K$ coincides the set of coherent states.
Let us take an optimal state $|\psi \rangle \in \mathcal K$. From the property of complementary channels and of the phase conjugation operator mentioned before we have that, $\mathcal A_\kappa(|\psi \rangle \langle \psi |)$, $\tilde{\mathcal A}_\kappa (|\psi \rangle \langle \psi |)$ and $T \circ \tilde{\mathcal A}_\kappa (|\psi \rangle \langle \psi |)$ have the same spectrum.  Since $F$ is unitary invariant, it necessarily depends only on the eigenvalues and we have
$F[\mathcal A_\kappa(|\psi \rangle \langle \psi |)]=F[\tilde{\mathcal A_\kappa}(|\psi \rangle \langle \psi |)]=F[T \circ  \tilde{\mathcal A_\kappa}(|\psi \rangle \langle \psi |)]$. Therefore,

\begin{equation}
F[\mathcal A_\kappa(|\psi \rangle \langle \psi |)]= F[\mathcal A_\kappa\circ  \mathcal E_\eta (|\psi \rangle \langle \psi |)]= F[ \sum_j p_j \mathcal A_\kappa ( |\psi_j \rangle \langle \psi_j | )],
\end{equation}
where $\{|\psi_j \rangle \}$ is the ensemble of states obtained after the beam splitter:

\begin{equation}\label{bs}
\mathcal E_\eta(|\psi \rangle \langle \psi |)= \sum_j p_j |\psi_j \rangle \langle \psi_j |.
\end{equation}
From the concavity of $F$ we have

\begin{equation}\label{concavity}
F[\mathcal A_\kappa (|\psi \rangle \langle \psi |)] \ge   \sum_j p_j
F[\mathcal A _\kappa ( |\psi_j \rangle \langle \psi_j | )].
\end{equation}
By hypothesis $|\psi \rangle  \in \mathcal K$ and so $F[\mathcal A_\kappa  (|\psi \rangle \langle \psi |)]\le F[\mathcal A_\kappa (|\psi_j \rangle \langle \psi_j |)] $ for each $j$. This can be
true only if the inequality (\ref{concavity}) is saturated and, from the hypothesis of strict concavity, we get

\begin{equation}
\mathcal A_\kappa ( |\psi_j \rangle \langle \psi_j | )=\rho_{out},\quad \forall j.
\end{equation}
From the definition of the quantum amplifier given in Eq.\ (\ref{amp}), it is evident that equal output states are possible only for equal input states: $ |\psi_j \rangle= |\psi'\rangle $ for every $j$. As a consequence
Eq.\ (\ref{bs}), reduces to
\begin{equation}\label{beamsp}
\mathcal E_\eta (|\psi \rangle \langle \psi |)= |\psi' \rangle \langle \psi' |.
\end{equation}
But now comes into play an important property of the beamsplitter which is known from the field of quantum optics \cite{beamsplitter1,beamsplitter2,beamsplitter3}, namely that
only coherent states remain pure under the action of a beamsplitter (Lemma 2 in section Methods).
Therefore, since Eq.\ (\ref{beamsp}) is valid for every choice of $|\psi\rangle \in \mathcal K$,  then $\mathcal K$ necessarily contains only coherent states. Moreover, for every Gaussian channel a displacement
of the input state corresponds to a (possibly different) displacement of the output state \cite{gauss2, gauss3}, which obviously does not change the entropy. Since coherent states are equivalent up to displacement operations  it means that $\mathcal K$ coincides with the whole set of coherent states.\\
\end{proof}

\noindent{\bf Majorization at the output of the channel.}
We can finally state our main result which proves the validity of the
majorization conjecture (B).
A graphical representation of the this property is given in Fig. \ref{noise}. \newline

\noindent \textbf{Proposition 1 } {Let $\Phi $ be a phase-insensitive
bosonic channel. Then, for every input state  $\rho $,
\begin{equation}
\Phi (|\alpha \rangle \langle \alpha |)\succ \Phi (\rho ),    \qquad  \forall \alpha \in \mathbb C,
\end{equation}%
where $|\alpha \rangle $ is any coherent state.}\newline

\begin{proof}  Let $\mathcal{F}$ be the class of real nonnegative strictly concave functions $f$ defined on
the segment $[0,1]$.
Consider the following functional
\begin{equation}\label{FUNC}
F(\rho )=\mathrm{Tr}f(\rho )=\sum_{j}f(\lambda _{j}^{\rho }),
\end{equation}
 where $f\in \mathcal{F}$. Then $F$ is well defined with values in $[0,+\infty]$,
since all the terms in the series are  nonnegative.
Moreover, it is unitary invariant and
and it can be shown that the strict
concavity of $f$ as a function of real numbers implies the strict concavity
of $F$ with respect to quantum states \cite{strict}. Therefore the
previous lemma can be applied and we get, for every state $\rho $ and every
strictly concave function $f$,
\begin{equation}
\sum_{j}f(\lambda _{j}^{\Phi (\rho )})\geq \sum_{j}f(\lambda _{j}^{\Phi
(|\alpha \rangle \langle \alpha |)}),\quad \forall \alpha \in \mathbb{C}.
\label{major}
\end{equation}
A well known theorem \cite{petz, concavity, nielsen} in the finite dimensional case
states that $\rho _{2}\succ \rho _{1}$ if and only if $\sum_{j}f(\lambda
_{j}^{\rho _{2}})\leq \sum_{j}f(\lambda _{j}^{\rho _{1}})$ for every concave
function $f$. Moreover, a similar result is valid also for strictly
concave functions $f\in \mathcal{F}$ and in infinite dimensions (see Lemma 3 in section Methods).  This concludes the proof.
\end{proof}\\

As a final remark, since the von Neumann entropy is a strictly concave
functional \cite{concavity}, we get an alternative proof (with respect to
the one given in \cite{vittorio}) of the minimal output entropy
conjecture. By applying Lemma 1 with the choice $F(\rho)=-\mathrm{Tr}[\rho
\log(\rho)]$, we get a slightly stronger version of the conjecture (A):
the minimum output entropy of a phase-insensitive channel is achieved
only by coherent input states.

Notice that, differently from the proof presented in Ref.~\cite{vittorio},
this result does not require the assumption that the mean energy of the
input should be finite and proves also that coherent states are the
unique optimizers.
Moreover, choosing  $f(x)=x-x^p$, $p>1$, leads to the proof of the similar statement for the minimal output Renyi entropies of all orders $p>1$.
\newline

\noindent{\bf \large Discussion}\\
 The main result of this paper is that every output state of a  phase-insensitive bosonic Gaussian channel is majorized by the output associated to a coherent input state (proof of the majorization conjecture).
We also prove that coherent input states are the unique minimizers of arbitrary nonnegative strictly concave output functionals and, in particular, of the von Neumann entropy (minimum output entropy conjecture). As compared to the proof of the minimal output entropy conjecture given in Ref.\ \cite{vittorio}, our result does not require the finiteness of the mean energy and proves the  uniqueness of coherent states.

Our work, while closing two longstanding open problems in quantum communication theory,  has a large variety of implications and consequences. For example, by using Lemma 1 and Proposition 1 one can:  compute the entanglement of formation of non-symmetric Gaussian states (see the last section of \cite{vittorio2}), evaluate the classical capacity of Gaussian channels \cite{vittorio2} and compute the exact quantum discord \cite{discord1} for a large class of channels \cite{discord2}. Moreover, from Proposition 1, we conclude that coherent input states minimize every Schur-concave output function like Renyi entropies of arbitrary order  \cite{evid1,evid3,konig1,konig2}. Finally, it is a simple implication that the pure entangled state $|\Psi_{out}\rangle$ obtained from a unitary dilation of
a  phase-insensitive Gaussian channel is more entangled than the output state $|\Psi_{out}\rangle'$ obtained with a coherent input. What is more, from the well known relationship between entanglement and majorization \cite{nielsen}, we also know that  $|\Psi_{out}\rangle'$ can be obtained from $|\Psi_{out}\rangle$ with local operations and classical communication.
The previous facts are just some important examples while a detailed analysis of all the possible implications will be the subject of future works.\\

\noindent{\bf \large Methods}\\
In order to make our analysis self-contained, in this section we present two properties (Lemma 1 and Lemma 2) which are used in the proof of the majorization conjecture.\\

\noindent \textbf{Lemma 2} {Coherent states are the only input pure
states which  produce a pure output for a beamsplitter. }\newline

This property is more or less implicit in several quantum optics papers \cite%
{beamsplitter1,beamsplitter2,beamsplitter3}. Here we present a complete proof, following an argument
similar to one used in Ref.\ \cite{beamsplitter1}, but using the formalism of
quantum characteristic functions.

\begin{proof}
Let $\mathcal{E}_\eta$ be the beamsplitter of transmissivity $\eta $, $0<\eta<1$, and
\begin{equation}
\mathcal{E_\eta}[|\psi \rangle \langle \psi |]=|\psi ^{\prime }\rangle
\langle \psi ^{\prime }|.
\end{equation}%
Then the complementary channel which is the beamsplitter of transmissivity $%
1-\eta $ satisfies a similar relation%
\begin{equation}
\mathcal{E}_{1-\eta}[|\psi \rangle \langle \psi |]=|\psi _{E}^{\prime
}\rangle \langle \psi _{E}^{\prime }|,
\end{equation}%
as the outputs of complementary channels have identical nonzero spectra.
Therefore we have
\begin{equation}
U(|\psi \rangle \otimes |0\rangle )=|\psi ^{\prime }\rangle \otimes |\psi
_{E}^{\prime }\rangle ,
\end{equation}%
where $U$ is the unitary implementing the minimal dilation of $\mathcal{E}_\eta.$
The corresponding canonical transformation of the annihilation operators $%
a,a_{E}$ for the system and the environment is
\begin{eqnarray}
U^{\dag }a U &=&\sqrt{\eta }a+\sqrt{(1-\eta )}a_{E} \\
U^{\dag }a_{E}U &=&\sqrt{(1-\eta )}a-\sqrt{\eta }a_{E},
\end{eqnarray}%
and the environment mode $a_{E}, a_{E}^{\dag}$ is in the vacuum state.
In phase space, this produces a symplectic transformation in the variables
of the characteristic functions:
\begin{equation}
\chi ^{\prime }(z)\chi ^{\prime }_{E}(z_{E})=\chi (\sqrt{\eta }z+\sqrt{(1-\eta )}%
z_{E})e^{\-\frac{1}{2}|\sqrt{(1-\eta )}z-\sqrt{\eta }z_{E}|^{2} }.
\label{FE}
\end{equation}%
By letting $z_{E}=0$ and $z=0$ respectively, we obtain%
\begin{eqnarray}
\chi ^{\prime }(z)&=&\chi (\sqrt{\eta }z)e^{ -\frac{1}{2}|\sqrt{(1-\eta )}%
z|^{2}} , \\
\quad \chi _{E}^{\prime }(z_{E})&=&\chi (\sqrt{(1-\eta )}z_{E}) e^{-\frac{1}{2}|\sqrt{%
\eta }z_{E}|^{2}}.
\end{eqnarray}%
Thus, after the change of variables $\sqrt{\eta }z\rightarrow z,\sqrt{%
(1-\eta )}z_{E}\rightarrow z_{E}$, and denoting $\omega (z)=\chi (z)e^{ \frac{1}{2}|z|^{2}} ,$ we get%
\begin{equation}
\omega (z)\omega (z_{E})=\omega (z+z_{E}).  \label{FE1}
\end{equation}%
The function $\omega (z),$ as well as the characteristic function $\chi (z),$
is continuous and satisfies $\omega (-z)=\overline{\omega (z)}.$ The only solution of
(\ref{FE1}) satisfying these conditions is the exponential function $\omega
(z)=\exp (\bar{z} \alpha -z \bar \alpha )$ for some complex $\alpha$. Thus we
obtain
\begin{equation}
\chi (z)=\exp \left[ \bar{z} \alpha -z \bar \alpha  -\frac{1}{2}|z|^{2}\right],
\end{equation}%
which is the characteristic function of a coherent state $|\alpha \rangle$.
\end{proof}\\

\noindent{\bf Lemma 3 }
{Given two (finite or infinite dimensional) vectors $\boldsymbol{\lambda}$ and $\boldsymbol \lambda'$ whose elements are nonnegative and normalized ($\sum_j \lambda_j=\sum_j \lambda'_j=1$),
 the following two relations are equivalent:
\begin{eqnarray}
 \boldsymbol{\lambda}'& \succ& \boldsymbol{\lambda} ,   \label{maj}\\
 \sum_j f (\lambda'_j) &\le& \sum_j f (\lambda_j) , \label{sumf}
\end{eqnarray}
for every function $f\in\mathcal{F}$,
where $\mathcal{F}$ is the class of real nonnegative strictly concave functions defined on
the segment $[0,1]$.} \\

\begin{proof}
It is well known in finite-dimensional majorization theory \cite{nielsen, petz,concavity,strict} that  if $\boldsymbol \lambda' \succ \boldsymbol \lambda$ then condition (\ref{sumf}) is satisfied for every concave function
 and so, in particular, for strictly concave functions. From the infinite dimensional generalization of the Horn-Schur theorem \cite{kaftal} one can extend this result to all functions $f\in\mathcal{F}$, using the fact that the series (\ref{sumf}) converges unconditionally to a value in $[0,+\infty]$.

 To prove the converse implication
suppose that the majorization relation (\ref{maj}) is not valid, then we construct an $f\in\mathcal{F}$ which violates the condition (\ref{sumf}).
As shown in Ref.\ \cite{concavity}, a simple concave (but non strictly concave) function can be found in a constructive way by using the following ansatz:
\begin{equation}
f^0(x):=\Bigg\{
 \begin{array}{l}
 \\
x,   \quad                   \mathrm{if }\;    0 \le x \le c,\\
\\
c,    \quad    \mathrm{if }\;    c \le x \le 1.\\
\\
 \end{array} \label{f0}
\end{equation}
If $\boldsymbol \lambda' \nsucc \boldsymbol \lambda$ then there exists a smallest integer $n$ for which $\sum_{j=1}^n \lambda'_{j} <\sum_{j=1}^n \lambda_j$. It is easy to show \cite{concavity} that,
by choosing $c=\lambda_n'$, the function $f^0$ violates the condition (\ref{sumf}), {\it i.e.} there is a positive and finite $\delta$ such that
\begin{equation} \label{delta}
\sum_j [ f^0(\lambda'_j)  -f^0 (\lambda_j)] =\delta >0.
\end{equation}
However, this does not conclude our proof because $f^0$ is not {strictly} concave. For this reason we take a slightly different function
\begin{equation}
 f^\epsilon(x):=f^0(x) - \epsilon x^2  ,
\end{equation}
which is strictly concave for every $\epsilon>0$ and belongs to the class $\mathcal{F}$.
Now, for an arbitrary vector $\boldsymbol \lambda$, by using the positivity and the normalization of the elements $\{\lambda_i\}$ we get the following convergence:
\begin{equation}
0 \le \sum_j [ f^0(\lambda_j) -f^\epsilon(\lambda_j)]\le  \epsilon .
\end{equation}
From the last continuity relation together with Eq.\ (\ref{delta}) we get:
\begin{eqnarray}
&&\sum_j[f^\epsilon(\lambda'_j)-f^\epsilon (\lambda_j) ] \ge \sum_j f^0(\lambda'_j)  - \sum_j f^0 (\lambda_j) -\epsilon \nonumber \\
&& =\delta -\epsilon.
\end{eqnarray}
The last term can be made positive by choosing $\epsilon < \delta$. Summarizing, we have shown that whenever the majorization relation (\ref{maj}) is not satisfied, there exists a small but finite $\epsilon$ such that $f^\epsilon$ violates the inequality (\ref{sumf}). Therefore the two conditions (\ref{maj}) and (\ref{sumf}) are equivalent.\\
\end{proof}

\vspace{1 em}

\noindent{\bf \large Acknowledgements}\\
The authors are grateful to R. F. Werner, J.
Oppenheim, A. Winter, L. Ambrosio, and M. E. Shirokov for comments and
discussions. AM acknowledges support from Progetto Giovani Ricercatori 2013 of SNS.
VG and AH also acknowledge support and catalysing role of the
Isaac Newton Institute for Mathematical Sciences, Cambridge, UK:
part of this work was conducted when attending the Newton Institute
programme \textit{Mathematical Challenges in Quantum Information}. AH
acknowledges the Rothschild Distinguished Visiting Fellowship which enabled
him to participate in the programme and partial support from RAS Fundamental
Research Programs, Russian Quantum Center and RFBR grant No 12-01-00319. \\

\noindent{\bf \large Author contributions}\\
All authors contributed equally to this work.\\

\noindent{\bf \large Competing financial interests}\\
The authors declare no competing financial interests.\\

\end{document}